\documentstyle[prl,aps,preprint,psfig]{revtex}
\begin{document}
\draft                                          
\title{
Spin resolved Andreev reflection in
ferromagnet-superconductor junctions with
Zeeman splitting}
\author{R\'egis M\'elin}
\address{Centre de Recherches sur les Tr\`es Basses
Temp\'eratures (CRTBT)\\
Laboratoire concentionn\'e avec l'Universit\'e Joseph
Fourier\\
CNRS, BP 166 X, 38042 Grenoble Cedex, France}
\date{\today}

\maketitle
\begin{abstract}
Andreev reflection in ferromagnet-superconductor junctions is
derived in a regime in which Zeeman splitting dominates the
response of the superconductor to an applied magnetic field.
Spin-up and spin-down Andreev reflections are shown
to be resolved as voltage is increased.
In the metallic limit, the
transition from Andreev to tunnel conductivity
in the spin-up channels has a non trivial
behavior when spin polarization is increased.
The conductance is asymmetric in a
voltage reversal.
\end{abstract}

The interplay between Andreev reflection
and spin polarization has generated recently
an important interest, both
theoretical~\cite{deJong,Falko,Jedema,Takahashi,Zutic}
and experimental~\cite{Petrashov2,Lawrence,Vasko,Giroud,Soulen,Upadhyay}.
The subgap conductance in normal
metal-superconductor (NS) junctions originates
from Andreev reflection~\cite{Andreev}: a spin-$\sigma$
electron incoming from the N side is reflected as a hole
in the spin-$(- \sigma)$ band while a spin-zero
Cooper pair is transferred into the superconductor.
Since the incoming electron and outgoing hole
belong to opposite spin bands, Andreev reflection
couples to a Fermi surface polarization in the
N side of the junction. de Jong and Beenakker~\cite{deJong}
showed theoretically that increasing the Fermi surface
polarization in ferromagnet-superconductor (FS) junctions
suppresses Andreev reflection because
Andreev reflection is limited 
by the minority-spin channels.
Their prediction was verified experimentally
by Soulen {\sl et al.}~\cite{Soulen} and
Upadhyay {\sl et al.}~\cite{Upadhyay},
who used this effect
to measure the Fermi surface polarization.
On the other hand, Tedrow and Meservey~\cite{Meservey}
demonstrated that under specific conditions, a
magnetic field can be used to tune a Zeeman
splitting of the quasiparticle excitations in a
superconductor~\cite{Meservey}, and used it
to perform a spin resolved tunnel spectroscopy
in FS junctions~\cite{Meservey}. I show in the
present Letter that Zeeman splitting 
can be used to resolve spin-up and
spin-down Andreev reflections,
with a different threshold voltage $e V_{\pm}
= \Delta \mp \mu_B H$ for the transition
from Andreev to tunnel conductivity
in a magnetic field $H$.
In NS junctions with Zeeman splitting,
the spin-up and spin-down
differential conductances have the same behavior at
the Andreev reflection threshold voltages $V_{\pm}$.
In FS junctions 
with Zeeman splitting,
a non trivial behavior at the spin-up threshold
voltage $V_+$ is predicted.
In addition, the conductance
is asymmetric in a voltage reversal.
These behaviors can be probed experimentally.

Our modeling neglects disorder in the superconductor,
as well as the proximity effect
in the N side of the
junction~\cite{Giroud,Hekking,Volkov,Petrashov,Courtois,Belzig,Gueron,Leadbeater}.
This approximation is justified in FS junctions,
where superconducting correlations do not extend
in the ferromagnet
beyond the exchange length $\sqrt{\hbar D / J}$
of order $20 \AA$~\cite{Giroud}.
In NS junctions, we expect the qualitative physics
arising from the coupling between
Andreev reflection and Zeeman splitting
to hold also in the presence of disorder.
Let us first consider a NS junction with Zeeman splitting.
The superconductor is assumed to have a thin
film geometry with the magnetic field
applied parallel to the film.
We assume a small orbital depairing parameter while
the critical field for destroying superconductivity is
set by Pauli paramagnetism~\cite{Clogston},
with large
values of $H_{c 2 \parallel} \sim 5 T$
for Al thin films~\cite{Meservey}.
The spin-orbit scattering length is supposed
to be small compared to the superconductor coherence length
$\xi$, as it is the case for light elements such
as Al~\cite{Meservey}. This insures that electrons
in the superconductor
have a well defined spin $\sigma$ at length $\xi$,
and therefore a well defined Zeeman energy
$- \mu_B H \sigma$~\cite{Meservey,Fulde}.
The coherence factors of spin-$\sigma$
electrons ($u_\sigma$) and holes
in the spin-($- \sigma$) band
($v_{- \sigma}$) with an energy $\epsilon$
are
\begin{equation}
\label{eq:coh}
u_{\sigma}^2 = 1 - v_{- \sigma}^2 =
{1 \over 2} \left( 1 + 
\frac{ \sqrt{ ( \epsilon + \sigma \mu_B H )^2
- | \Delta |^2 }}
{ \epsilon + \sigma \mu_B H } \right)
,
\end{equation}
with therefore a coupling between
Andreev reflection and Zeeman splitting.
A step function variation of the superconducting
gap at the interface is assumed: $\Delta(x) = \Delta
\theta(x)$. We consider a $\delta$-function
elastic interface
scattering potential $V(x) = H_0 \delta(x)$,
interpolating between a
metallic contact if $H_0=0$ and a
tunnel junction if $H_0 = \infty$~\cite{BTK}.
The interface
barrier is normalized with respect to the
Fermi velocity: $Z = m H_0/
(\hbar \sqrt{2 m \mu})$, with
$\mu=\hbar^2 k_F^2 / 2 m$ the
chemical potential~\cite{BTK}.
The energy dependence of the transmitted
quasiparticle wave vectors is irrelevant
to the present calculation~\cite{note}
and we consider identical Fermi
wave vectors in the  superconductor and the
normal metal since this assumption does not
change the qualitative physics.
Given the coherence factors (\ref{eq:coh}),
the Andreev reflection transition probability
of electrons with a spin-up
and holes in the spin-down band
with an energy $\epsilon$ is
\begin{equation}
\label{eq:Aeup}
A^{e \uparrow}(\epsilon) =
A^{h \downarrow}(\epsilon) =
A_{\rm BTK} (\epsilon + \mu_B H)
,
\end{equation}
with $A_{\rm BTK}(\epsilon)$ the Blonder, Thinkham and
Klapwijk (BTK) Andreev reflection coefficient~\cite{BTK}.
Similarly in the spin-down sector,
\begin{equation}
\label{eq:Aedown}
A^{e \downarrow}(\epsilon) =
A^{h \uparrow}(\epsilon) =
A_{\rm BTK} (\epsilon - \mu_B H)
.
\end{equation}
Eqs.~\ref{eq:Aeup} and~\ref{eq:Aedown} are valid
also if $\epsilon < 0$, in which case transmission
of quasiparticles on negative energy branches should
be considered. Noting $B_{\rm BTK}$ the
BTK backscattering coefficient~\cite{BTK}, the
zero-temperature differential conductance
of spin-$\sigma$ carriers
\begin{equation}
\label{eq:dIdV}
{ d I^\sigma \over d V}(e V,H) =
{e^2 \over h} \left[
1 + A_{\rm BTK}(e V + \sigma \mu_B H) -
B_{\rm BTK}(eV + \sigma \mu_B H)
\right]
\end{equation}
shows a Zeeman splitting for an arbitrary
interface scattering in the sense that the
magnetic field enters the conductivity
{\sl via} the combination $eV + \sigma \mu_B H$
only. The tunnel spectrum in the limit $Z \gg 1$
reproduces the Zeeman splitted density of
states of the superconductor $\rho_\sigma
( \epsilon ) = \rho_{\rm BCS}(\epsilon
+ \sigma \mu_B H)$, with $\rho_{\rm BCS}$
the single-spin BCS density of states~\cite{Meservey,Fulde,BTK}.
In the metallic limit $Z=0$
and below the spin-up threshold voltage
$e V_+  = \Delta - \mu_B H$,
spin-up and spin-down transport originate
from Andreev reflection, with a conductance
of $2 e^2 / h$ per spin channel
(see Fig.~\ref{fig1}).
Spin-up transport transits from
Andreev reflection to tunneling
at the spin-up threshold voltage,
smaller than the
spin-down threshold voltage
$e V_- = \Delta + \mu_B H$.
In between $V_+$ and $V_-$ a
plateau of $3 e^2/(2h)$ per spin channel
develops in the conductance when $H$
increases, corresponding to an Andreev
reflection transport of
spin-down carriers and a tunnel
transport of spin-up carriers.
Notice that the single-spin
conductance in Eq.~\ref{eq:dIdV}
is not symmetric in a voltage reversal.
The total conductance
of the NS junction is however
symmetric in a voltage reversal because
the two spin channels play a symmetric role
in this junction.

We now extend our treatment to incorporate
the effect of a spin polarization in the
normal metal.
We show a non trivial transition
from Andreev to tunnel transport
at the spin-up threshold voltage
$V_+$, as
well as a conductance asymmetric in a voltage
reversal. We denote by $n$
and $n'$ the quantum numbers
associated to a quantized transverse
motion in a clean FS point
contact of cross sectional area $a^2$.
We assume a Stoner ferromagnet
with an exchange field
$h_{\rm ex}(x) = h_{\rm ex} \theta(-x)$.
The channel with
transverse quantum numbers $(n,n')$
in the spin-$\sigma$ band 
has a dispersion
$$
E_{n,n'}^\sigma (k^\sigma) =
\frac{ \hbar^2 (k^\sigma)^2 }
{2 m} - \sigma h_{\rm ex} + \kappa (n^2
+ n'^2)
,
$$
with the energy
$\kappa= (\hbar^2 /2 m) (\pi / a)^2$
inverse proportional to the junction
area, and related to the number of
spin-$\sigma$ channels according to
$N^\sigma = \pi (\mu + \sigma h_{\rm ex})
/ (4 \kappa)$~\cite{Upadhyay}.
The associated barrier parameter
$Z^{F,\sigma}_{n,n'}$ of
spin-$\sigma$ electrons in the
channel $(n,n')$ is
$
Z^{F,\sigma}_{n,n'}
= \left(
1 + \sigma {h_{\rm ex} \over \mu}
- {\kappa \over \mu}
(n^2 + n'^2) \right)^{-1/2}
Z ,
$
with $Z = m H_0 /( \hbar \sqrt{2 m \mu})$.
The transverse dimensions of the S
side of the junction are assumed
to be identical to the ones of the N side
and the gap, the interface scattering
and the exchange field
are constant in the transverse direction,
with therefore a conservation of the
transverse quantum numbers
across the interface\cite{note-bis}.
The pairing Hamiltonian in the S side 
with a cross sectional area $a^2$ is
$$
H^S = \sum_{n,n',k,\sigma} \left(
\frac{ \hbar^2 k^2 }{2 m}
+ \kappa (n^2+n'^2) \right) c_{n,n',k,\sigma}^+
c_{n,n',k,\sigma}
+ \sum_{n,n',k} \left( \Delta
c_{n,n',k,\uparrow}^+ c_{n,n',-k,\downarrow}^+
+ h.c. \right)
,
$$
with an associated barrier parameter
$Z^S_{n,n'} = \left( 1 - \frac{\kappa}{\mu}
(n^2 + n'^2) \right)^{-1/2} Z$ 
different from
$Z^{F,\sigma}_{n,n'}$ because of the
Fermi wave vector mismatch
between the ferromagnet and
the superconductor~\cite{Zutic}.
The channels
with a spin-up Fermi surface only have
a real positive $Z^{F,\uparrow}_{n,n'}$
and a pure imaginary $Z^{F,
\downarrow}_{n,n'}$. Physically, 
a spin-up electron incoming from
the N side
below the superconducting gap in such a channel
is Andreev reflected into an evanescent
hole state in the spin-down band, with
a pure imaginary wave vector $k^\downarrow$.
The hole propagates in the ferromagnet
over the length scale $1/{\rm Im}
(k^\downarrow)$ before it is
backscattered onto the interface
and Andreev reflected as a spin-up electron,
therefore not carrying current,
as proposed by de Jong and Beenakker~\cite{deJong}.
Incorporating this process under the form of
a pure imaginary interface scattering
allows to calculate transport above
the superconducting gap.
The matching of the
wave functions between the F and S sides
is solved similarly to Ref.~\cite{Zutic},
including the coherence factors in Eq.~\ref{eq:coh},
and the barrier parameters $Z^{F,\sigma}_{n,n'}$
and $Z^S_{n,n'}$~\cite{note-ter}.
The differential conductance spectra
are shown on Fig.~\ref{fig2} in the metallic limit
$Z=0$.  At low voltage, the conductance
shows a reduction of Andreev reflection
by spin polarization~\cite{deJong}.
The large voltage limiting value
of the tunnel conductance per spin channel
decreases from the Landauer
value $e^2 / h$ 
in the absence of spin
polarization to $e^2 / (2 h)$ with a full
polarization, because
only the ferromagnet
channels with a corresponding channel in
the superconductor contribute to the tunnel
conductance. The number of spin-down
tunneling channels is $N^\downarrow$,
while spin-up tunneling
is limited by the number of
superconducting channels 
$N^S = \pi \mu / 4 \kappa$.
The total number of tunneling channels
is $\pi (2 \mu - h_{\rm ex})/4 \kappa$,
reduced by a factor of two when 
the exchange field $h_{\rm ex}$
increases from zero to $\mu$.

Now the behavior of the differential 
conductance at the spin-up threshold
voltage $e V_+ = \Delta - \mu_B
H$ differs qualitatively in the
weak and strong polarization
regimes:  the conductance
decreases with voltage at $e V_+$
if spin polarization is weak while
it increases if spin polarization
is strong (see Fig.~\ref{fig2}).
With a weak polarization,
most of the spin-up channels
are Andreev reflected and the decrease in
conductance at $e V_+$ can be understood
qualitatively
on the basis of the transition from
Andreev to tunnel transport in the
single channel BTK model~\cite{BTK}.
If spin polarization is strong,
a fraction $1-(N^\downarrow / N^\uparrow)$ of
the spin-up channels are not Andreev reflected
if $V < V_+$. These channels however
contribute to the tunnel current if
$V > V_+$, with a spin-up tunnel
conductance $\simeq (e^2 / h) N^S$,
larger than the Andreev conductance
$\simeq (2 e^2 / h) N^\downarrow$
if $h_{\rm ex} > \mu/2$.
The conductance
at the spin-down threshold voltage
behaves similarly to
the single channel BTK model
because there is no suppression
of Andreev reflection in the spin-down channels.
As a result of the different behavior in the
spin-up and spin-down channels,
the conductance spectrum is
asymmetric in a voltage reversal
(see Fig.~\ref{fig2}). 
Tedrow and
Meservey used this asymmetry to
probe spin polarization in the tunnel
limit~\cite{Meservey}.
We predict an asymmetry in the
metallic limit also, which
can be used as a signature
of the present effect in an experiment.

Finally, we have shown on Fig.~\ref{fig3}
the behavior of the FS junction
model with an interface scattering potential
equal to the Fermi velocity ($Z=1$).
In this parameter range, and in the
absence of spin polarization,
two tunnel-like peaks coexist with a
finite low-voltage conductance
originating from Andreev reflection.
The subgap conductance of the unpolarized
junction is smaller than $2 e^2 / h$ because of 
the finite interface scattering.
Increasing spin polarization results
in a suppression
of Andreev reflection by spin
polarization and
spin polarized tunneling (a spin-up
peak at $e V_+$ with a stronger weight
than the spin-down peak at $e V_-$).
These two phenomena may therefore be
observed simultaneously.

To conclude, we have shown that Zeeman
splitting can be used to resolve
the spin-up and spin-down Andreev reflections
in NS and FS junctions. In metallic FS junctions,
the spin-up tunnel current is larger
than the spin-up Andreev reflection current
if spin polarization is large, while
the transition from Andreev
to tunnel transport in the spin-down channels
has a BTK-type behavior.
The different behavior in
the spin-up and spin-down channels generates
a conductance spectrum asymmetric in a voltage
reversal.

I would like to acknowledge fruitful
discussions with O.~Bourgeois, 
H.~Courtois,
M.~Devoret, D.~Est\`eve,
D.~Feinberg, P.~Gandit, M.~Giroud, F.~Hekking
and B.~Pannetier.

\begin{figure}[tbp] 
\caption{Differential conductance of the NS junction in
the metallic limit $Z=0$, with a Zeeman splitting
$\mu_B H = 0$~($\Diamond$),
$0.2$~($+$), $0.4$~($\Box$),
$0.6$~($\times$) in units of
the superconducting gap $\Delta$.
The conductance is
normalized to the number of spin channels.
The voltage is measured in units of the
superconducting gap $\Delta$. The conductance
is symmetric in a voltage reversal. A plateau
of $3 e^2 / (2 h)$ per spin channel develops
in the conductance when $H$ increases.
}
\label{fig1} 
\end{figure}

\begin{figure}[tbp] 
\caption{Differential conductance of the FS junction in
the metallic limit $Z=0$. The conductance is
normalized to the number of spin channels.
The voltage is measured in units of the
superconducting gap $\Delta$. The chemical
potential is $\mu = 10^4$, and the Zeeman
splitting is $\mu_B H = 0.3$ (in units
of $\Delta$). The Fermi surface polarization
$P=(N^\uparrow - N^\downarrow) / (N^\uparrow
+ N^\downarrow)$ are
$P=0$~($\Diamond$), $P=0.21$~($+$),
$P=0.42$~($\Box$), $P=0.63$~($\times$),
and $P=0.83$~($\bigtriangleup$).
The conductance is asymmetric in a voltage reversal,
and has a non trivial behavior at the spin-up
threshold voltage $e V_+ = \Delta - \mu_B H$.
}
\label{fig2} 
\end{figure}

\begin{figure}[tbp] 
\caption{Differential conductance of the FS junction with
$Z=1$. The conductance is
normalized to the number of spin channels.
The voltage is measured in units of the
superconducting gap $\Delta$. The chemical
potential is $\mu = 10^4$, and the Zeeman
splitting is $\mu_B H = 0.3$ (in units
of $\Delta$). The Fermi surface polarization
$P=(N^\uparrow - N^\downarrow) / (N^\uparrow
+ N^\downarrow)$ are
$P=0$~($\Diamond$), and $P=0.83$~($\Box$).
Spin polarization results simultaneously
in a suppression
of Andreev reflection and spin polarized
tunneling.
}
\label{fig3} 
\end{figure}

\newpage

\centerline{\psfig{figure=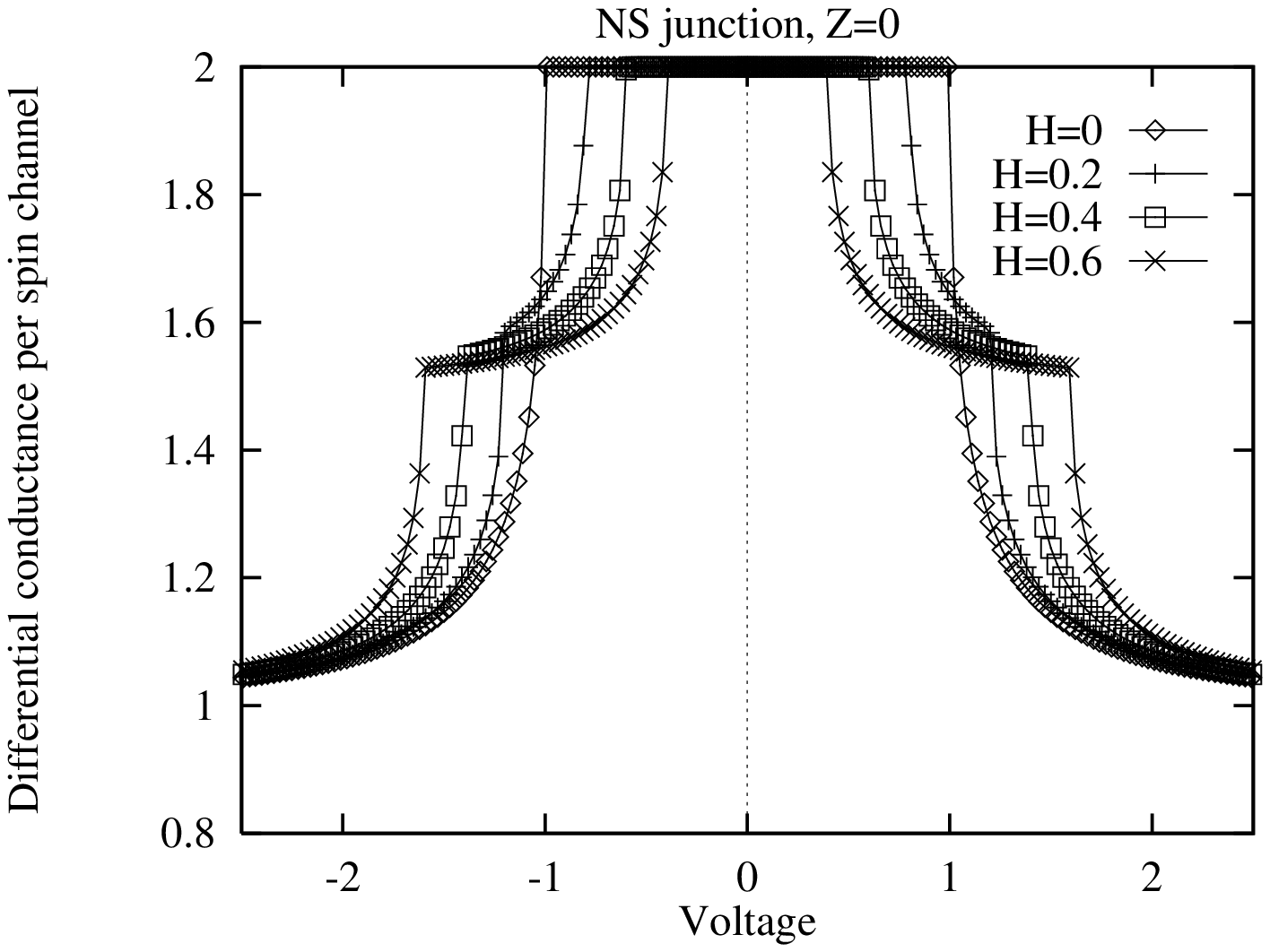,width=15cm,height=10cm}}

\newpage

\centerline{\psfig{figure=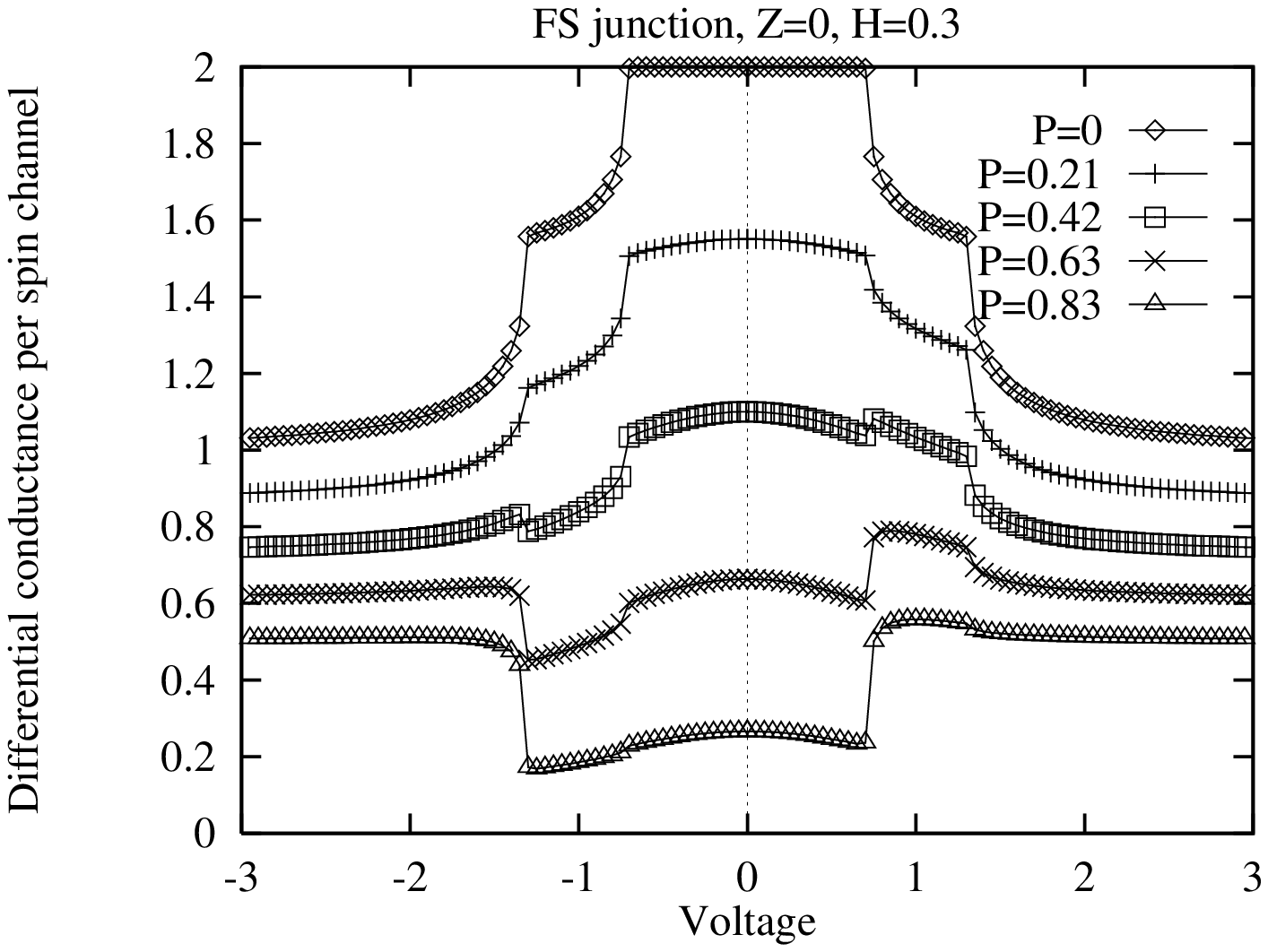,width=15cm,height=10cm}}

\newpage

\centerline{\psfig{figure=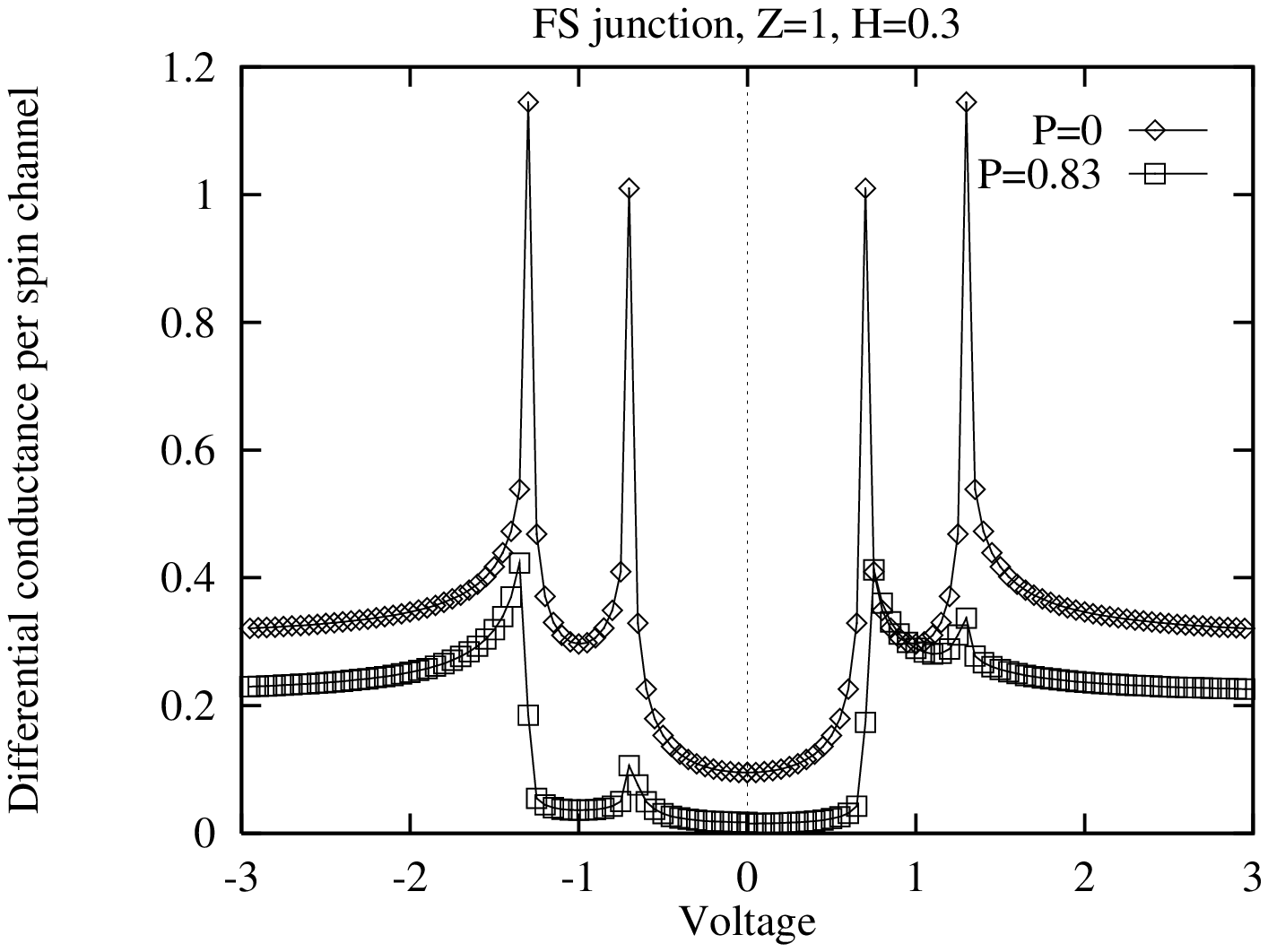,width=15cm,height=10cm}}


\begin{references}
\bibitem{note} We calculated the Andreev
reflection coefficient of NS junctions
with Zeeman splitting
with the full dependence
of the quasiparticle wave vectors on $H$. 
We neglect
the resulting correction of order
$\mu_B H / \mu \ll 1$
to the Andreev reflection coefficient.
\bibitem{note-ter}  In the energy range
$-\Delta + \mu_B H < \epsilon < \Delta - \mu_B H$,
the backscattering transition probability is
unity, with a real
$Z_{n,n'}^{F,\uparrow}$ and a pure imaginary
$Z_{n,n'}^{F,\downarrow}$ while $Z$ is
arbitrary. This shows the total backscattering
of spin-up electrons incoming below the superconducting
gap in a channel with no spin-down Fermi surface.
\end{references}
\end{document}